\documentclass[aps,prd,showpacs,twocolumn,floatfix]{revtex4}

\usepackage{graphicx}% Include figure files
\usepackage{dcolumn}% Align table columns on decimal point
\usepackage{bm}% bold math

\begin{document}

\def\be{\begin{equation}}
\def\ee{\end{equation}}
\def\bd{\begin{displaymath}}
\def\ed{\end{displaymath}}
\def\F{\rm F}
\def\J{\rm J}
\def\L{\rm L}
\def\M{\rm M}
\def\N{\rm N}
\def\P{\rm P}
\def\S{\rm S}
\def\V{\rm V}

\title{\Large Nucleon-Meson Coupling Constants and Form Factors\\
in the Quark Model}
\author{
C.Downum,$^a$\footnote{Email: c.downum1@physics.ox.ac.uk}
T.Barnes,$^{b,c}$\footnote{Email: tbarnes@utk.edu}
J.R.Stone$^{a}$\footnote{Email: j.stone1@physics.ox.ac.uk}
and
E.S.Swanson$^d$\footnote{Email: swansone@pitt.edu}
}
\affiliation{
$^a$Department of Physics,
Oxford University, Oxford, OX1 3PU, UK\\
$^b$Physics Division, Oak Ridge National Laboratory,
Oak Ridge, TN 37831, USA\\
$^c$Department of Physics and Astronomy, 
University of Tennessee,
Knoxville, TN 37996,
USA\\
$^d$Department of Physics and Astronomy, 
University of Pittsburgh, 
Pittsburgh, PA 15260, 
USA\\
}

%\date{\today}

\begin{abstract}
We demonstrate the calculation of the coupling constants and form factors 
required by effective hadron lagrangians using the quark model. 
These relations follow from equating expressions for strong transition 
amplitudes in the two approaches. 
As examples we derive the NN$m$ nucleon-meson coupling constants 
and form factors for $m = \pi, \eta, \eta\, ', \sigma, a_0, \omega$ and $\rho$, 
using harmonic oscillator quark model meson and baryon wavefunctions 
and the $^3$P$_0$ decay model; this is a first step towards deriving a quark-based
model of the NN force at all separations. This technique should be useful 
in the application of effective lagrangians to processes in which 
the lack of data precludes the direct determination of coupling constants 
and form factors from experiment.

\end{abstract}

\maketitle

\section{Introduction}

Effective hadronic lagrangians are widely used 
the description of the interactions of hadrons. In this method a distinct
quantum field is introduced for each relevant hadronic species, and 
interactions are assumed for these fields that are consistent with 
known symmetries and conservation laws. Although a more physically justified
description of hadron scattering amplitudes would employ momentum-dependent 
coupling constants (form factors), it often suffices near threshold to assume 
pointlike ``hard" hadronic vertices with fixed coupling constants. 
This approximation may be relaxed by incorporating hadronic form factors 
as a power series of gradient interactions. 
Unfortunately, this typically leads to nonrenormalizable 
ultraviolet divergences, which requires the determination 
of many coupling constants from experiment. 

When using effective lagrangians one typically ignores 
the existence of hadronic substructure 
(quarks and gluons), and determines each coupling constant in the effective 
lagrangian from experiment. Although this procedure is feasible 
in experimentally highly constrained problems such as the 
NN interaction, in other processes with little data  
the coupling constants are simply assigned plausible values. 
In such cases it would be very useful to have numerical 
estimates of the coupling constants. Since these coupling constants and
form factors are determined by the underlying QCD degrees of freedom, 
one may evaluate them directly in terms of the interactions of quarks 
and gluons and the substructure of the hadronic bound states. 

In this letter we will investigate this relation between 
effective lagrangian couplings and quark model bound state hadron
wavefunctions. The specific cases we consider are the meson-nucleon 
couplings, which are usually fitted to data in meson exchange models 
of the NN force. These NN$m$ vertices are chosen as our initial examples 
in part because they are among the most important hadronic couplings 
for nuclear physics applications, and also because the NN$\pi$ coupling 
constant $g_{\N\N\pi} \approx 13.5$ is the best determined strong 
coupling in hadron physics. It is clearly of interest to determine 
whether this experimental coupling constant is consistent with the 
value predicted using quark model hadron wavefunctions. 

We regard this computation as the first step in an attempt to compute the 
strong interactions of nucleons at all separations from microscopic quark-based models. 
In this approach the short ranged NN interactions are dominated by the
quark-gluon forces encountered by overlapping nucleon quark wavefunctions, 
and the long ranged NN interactions are mediated by $t$-channel meson exchange. 
The meson-nucleon couplings in $t$-channel meson exchange, which we compute here, 
are themselves determined by the overlap of meson and nucleon quark wavefunctions. 

\vfill\eject

\section{Technique}

Our method for defining effective hadronic coupling constants and form factors 
is to equate specific hadron emission amplitudes 
predicted by the effective hadron lagrangian to the corresponding 
decay amplitude in the quark model. 
Applied to NN$m$ couplings in particular, we require that
\be
\langle Nm |{H_{eff}}| N \rangle = 
\langle N(q^3)m(q\bar q)  | H_{decay}| N(q^3) \rangle \ .
\label{eq:H_identity}
\ee  
This general approach was used previously by the Orsay group of
LeYaouanc {\it et al.} \cite{LeYaouanc:1972ae}, 
who considered $\rho\pi\pi$, NN$\pi$ and NN$\rho$ couplings; 
the relation between this earlier reference and our results will be discussed.
A similar quark model approach was more recently applied to the determination
of HQET electroweak form factors by Isgur {\it et al.} \cite{Isgur:1988gb}.

Ideally one would impose this relation using a relativistic quark model, 
in which case there would be no difficulty in identifying the
effective lagrangian matrix element with the quark model result.
As the nonrelativistic quark model formalism is much better
established, we will instead use nonrelativistic quark model wavefunctions,
and apply our defining relation Eq.(\ref{eq:H_identity}) 
between matrix elements near threshold. 
Since the quark model wavefunctions we use are nonrelativistic, 
the form factors we extract by equating quark model and (relativistic) 
field theory matrix elements have an ambiguity in how we relate
results in different frames. 
In this initial study we will simply 
assume a particular frame, the ``decay frame" of a rest initial hadron,
as in the earlier work of the Orsay group \cite{LeYaouanc:1972ae}.
The complication of implicit frame dependence will be investigated
in detail a subsequent study. 

It is important to note that these frame ambiguities are resolved 
in the weak binding and nonrelativistic limits, in which the
quark model states approach the transformation properties specified by
the Lorentz group. Thus our quark model predictions of form factors 
are best justified at zero three-momentum, where we define the coupling 
constants. Of course the model assumption is that the predicted form factors 
remain useful estimates at moderate nonzero momentum.

The specific quark model strong decay Hamiltonian we use is the 
$^3\P_0$ model \cite{LeYaouanc:1972ae,Micu:1968mk,Ackleh:1996yt},
which has seen very wide application, and is known
to give reasonably accurate numerical results for both meson
\cite{Barnes:1996ff,Barnes:2002mu,Barnes:2005pb} and 
baryon \cite{Capstick:1993kb} strong decays. 

\subsection{NN$\pi$, NN$\eta$ and NN$\eta'$}

As a first example we consider the NN$\pi$ coupling. We assume the standard
pseudoscalar effective lagrangian
\be
{\cal L}_{\N\N\pi} = -ig_{\N\N\pi} \bar \Psi \gamma_5 \vec \tau\,\Psi 
\cdot \vec \phi_{\pi}\ ,
\ee
where $\Psi = [\psi_p, \psi_n]$ is an isodoublet of Dirac nucleon fields,
the $\vec \tau$ are isospin Pauli matrices, and $\vec \phi_{\pi}$ 
is an isotriplet of pion fields, 
$[\, \phi_{\pi^+},\phi_{\pi^0},\phi_{\pi^-}]$. Specializing to the transition
$p(+1/2) \to p(+1/2)\; \pi^0 $ (an initial spin-up proton at rest going to
a spin-up proton with momentum $-\vec {\P}$ and a recoiling $\pi^0$ with momentum
$+ \vec {\P}$), we find the matrix element 
\bd
\langle p\pi^0  | H_{eff}| p \rangle / 
\delta(\vec {\P}_f - \vec {\P}_i) \equiv h_{fi} = 
\ed
\be
i g_{\N\N\pi}\, \frac{M_p}{E_p}\, \frac{1}{\sqrt{(2\pi)^3 2E_\pi}}\,
\Big[ \frac{\P\cos(\theta)}{\sqrt{2M_p(E_p+M_p)}}\Big].
\ee
We similarly evaluate the quark model matrix element of the $^3\P_0$ decay
model Hamiltonian for $p(+1/2) \to p(+1/2)\; \pi^0 $,
using the techniques and Gaussian baryon and meson 
wavefunctions given in Refs.\cite{Ackleh:1996yt,Barnes:1993nu}.
The result is

\bd
\langle p(q^3)\pi^0(q\bar q)  | H_{decay}| p(q^3) \rangle / 
\delta(\vec {\P}_f - \vec {\P}_i) \equiv h_{fi} = 
\ed
\bd
-\, \gamma\, \frac{10}{9\pi^{3/4}}\frac{(1+r^2/4)\, r^{3/2}}{(1+r^2/3)^{5/2}}
\; \frac{{\P}\cos(\theta)}{\alpha^{3/2}} \; \cdot 
\ed
\be 
\exp\bigg\{ -\frac{1}{6}\frac{(1+5r^2/12)}{(1+r^2/3)}\; 
\bigg( \frac{\P}{\alpha}\bigg)^2 \bigg\}.
\label{eq:NNpi_FF}
\ee
where $\alpha$ and $\beta = \alpha/r $ are the baryon and meson 
Gaussian quark wavefunction inverse length scales,
and $\gamma$ is the dimensionless $^3\P_0$ model $q\bar q$ pair production amplitude. 
This result is consistent with the earlier result of  
LeYaouanc {\it et al.} \cite{LeYaouanc:1972ae}, given the parameter relations
between that reference and the current work,
$(R_{\N} = 1/\alpha)$, $(R_\pi = 1/\beta)$ and 
$(\gamma_{LeY} = \sqrt{24\pi}\gamma)$. 
Typical values for these parameters found in the literature are
$\alpha = 0.25-0.4$~GeV, $\beta = 0.3-0.4$~GeV, 
and $\gamma = 0.4-0.5$. Here we fix
$\gamma = 0.4$ (taken from our extensive studies of light and charmed meson
strong decays \cite{Ackleh:1996yt,Barnes:1996ff,Barnes:2002mu,Barnes:2005pb}),
and show numerical results for this range of 
$\alpha$ and $\beta$.

\begin{figure}[ht]
\vskip 0.7cm
\includegraphics[width=8.5cm]{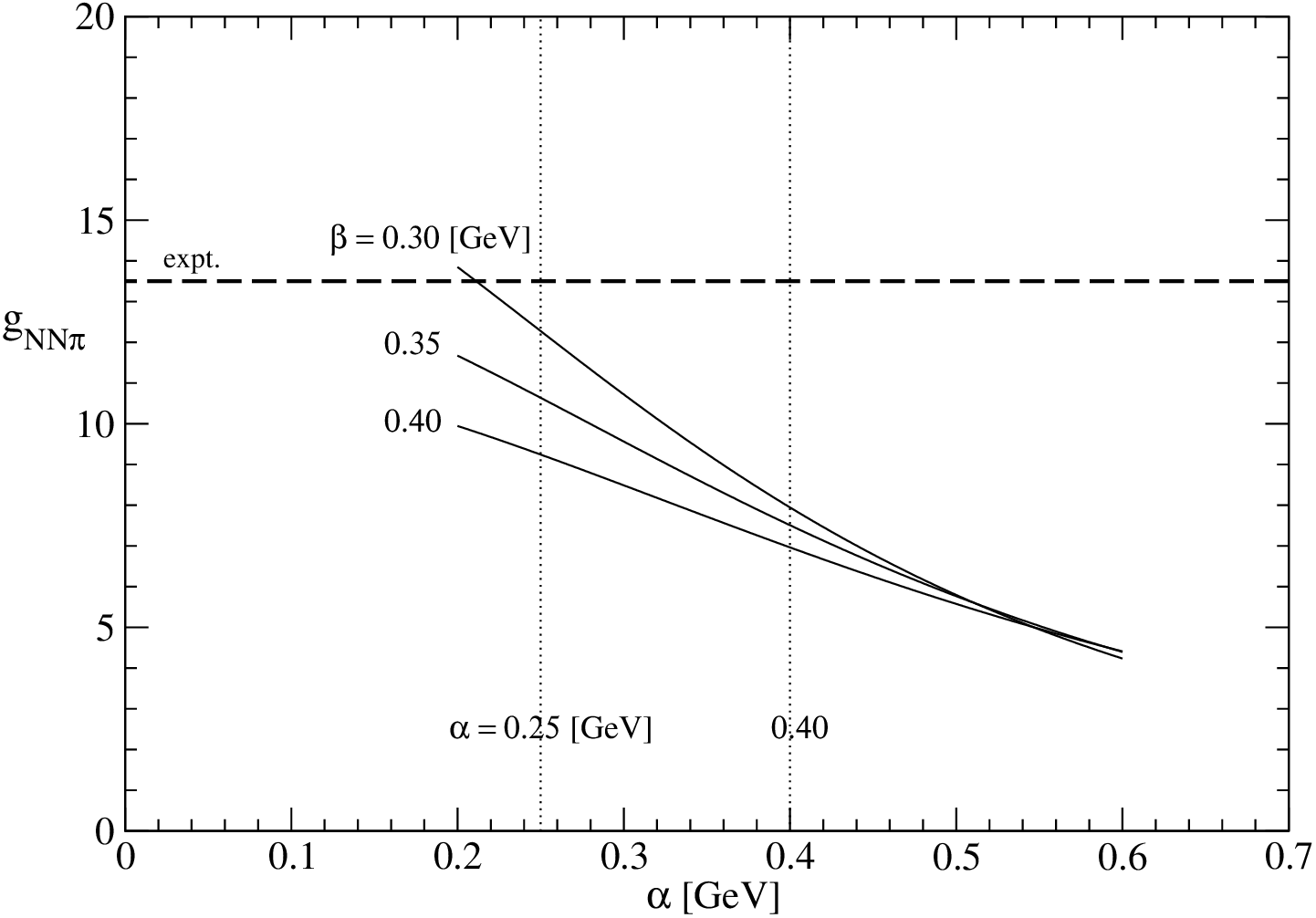}
\caption{
The theoretical pion-nucleon coupling constant 
$g_{\N\N\pi}$ 
(Eq.\ref{eq:g_NNpi}) 
as a function of the quark model Gaussian wavefunction
length scales $\alpha$ (baryons) and $\beta$ (mesons). The range
of values of $\alpha$ and $\beta$ typically found in the quark model 
literature is indicated (see text). The $^3$P$_0$ model $q\bar q$ 
pair production amplitude $\gamma = 0.4$ was taken from studies of meson decays.
The experimental $g_{\N\N\pi} \approx 13.5$ is also shown.}
\label{fig:g_NNpi}
\end{figure}

The NN$\pi$ coupling constant we find by equating these expressions at 
threshold is
\be
g_{\N\N\pi} =  \gamma \, \frac{2^4\cdot 5}{3^2}\, \pi^{3/4}\,  
\frac{(1+r^2/4)\, r^{3/2}}{(1+r^2/3)^{5/2}}\;  
\frac{M_p m_{\pi}^{1/2} }{\alpha^{3/2} }.
\label{eq:g_NNpi}
\ee
(We have suppressed an overall phase factor of $+i$, which is
normalization convention dependent.)
The numerical value of this coupling constant as a function of
$\alpha$ and $\beta$ is shown in Fig.\ref{fig:g_NNpi}. 
Evidently the range of typical wavefunction length scale parameters
$\alpha$ and $\beta$ leads to a factor of two variation in the theoretical 
$g_{\N\N\pi}$; it ranges between $7.0-12.2$ for the parameters shown 
in the figure. The experimental value of $\approx 13.5$ evidently 
corresponds to values of $\alpha$ and $\beta$ near the lower end 
of their respective ranges, provided that we fix the $q\bar q$ pair 
production amplitude at the meson decay value of $\gamma = 0.4$.

\begin{figure}[ht]
\vskip 0.7cm
\includegraphics[width=8.5cm]{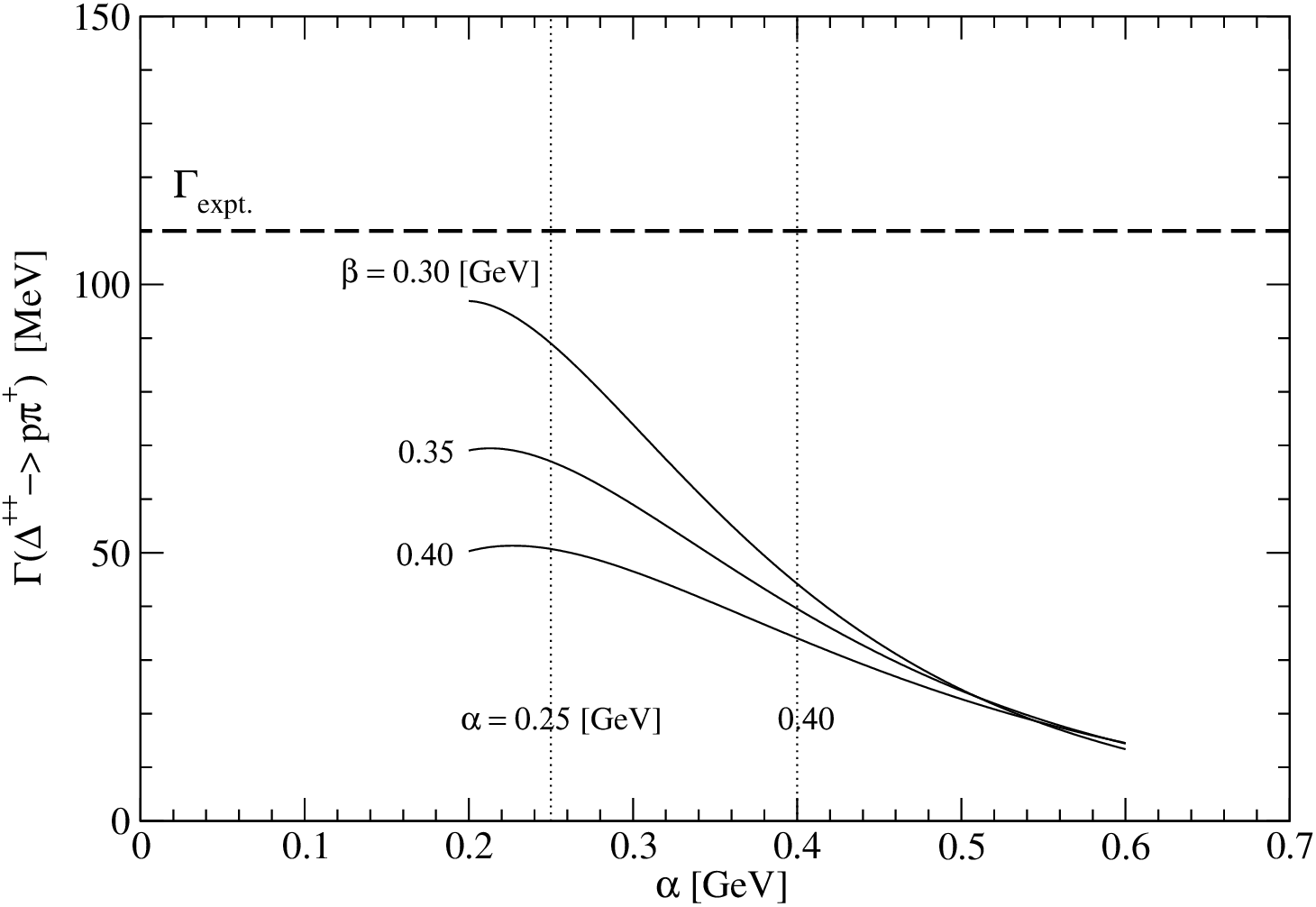}
\caption{
The theoretical $^3$P$_0$ quark model $\Gamma(\Delta^{++} \to p \pi^+)$
width (Eq.\ref{eq:DNpi}) 
as a function of the wavefunction length scales $\alpha$ and $\beta$, for
a pair production amplitude of $\gamma = 0.4$. 
The experimental width is also shown.}
\label{fig:DNpi}
\end{figure}

Since the ``experimental" value of $g_{\N\N\pi}$ is  
not actually based on a direct observation of pion emission, it is prudent 
to carry out an independent calculation of a closely related decay 
process that does involve a detected pion in the final state. 
The transitions 
$\Delta \to \N \pi$ are useful in this regard because the matrix
elements are related to the NN$\pi$ coupling by SU(6) symmetry, 
assuming identical spatial wavefunctions. Specializing to 
$\Delta^{++} \to p \pi^+$, the quark model result for the total 
width is
\bd
\Gamma(\Delta^{++} \to p \pi^+) = 
\gamma^2 
{\pi}^{1/2}
\frac{2^8}{3^3}
r^3 \frac{(1+r^2/4)^2}{(1+r^2/3)^5}\, \cdot
\ed
\be
\frac{E_pE_\pi}{M_\Delta}  
\bigg( \frac{\P}{\alpha} \bigg)^3
\exp\bigg\{
{-\frac{1}{3} \frac{(1+5r^2/12)}{(1+r^2/3)} \bigg( \frac{\P}{\alpha} \bigg)^2 }
\bigg\}
\label{eq:DNpi}
\ee
This theoretical quark model decay rate is shown in Fig.\ref{fig:DNpi} 
for the same range of wavefunction parameters $\alpha,\beta$ as 
$g_{\N\N\pi}$ in Fig.\ref{fig:g_NNpi}, and the experimental width 
of 110~MeV is also shown.
Evidently the parameter constraints due to $g_{\N\N\pi}$
and $\Gamma(\Delta \to \N \pi)$ are approximately consistent. 

\begin{figure}[ht]
\vskip 0.7cm
\includegraphics[width=8.5cm]{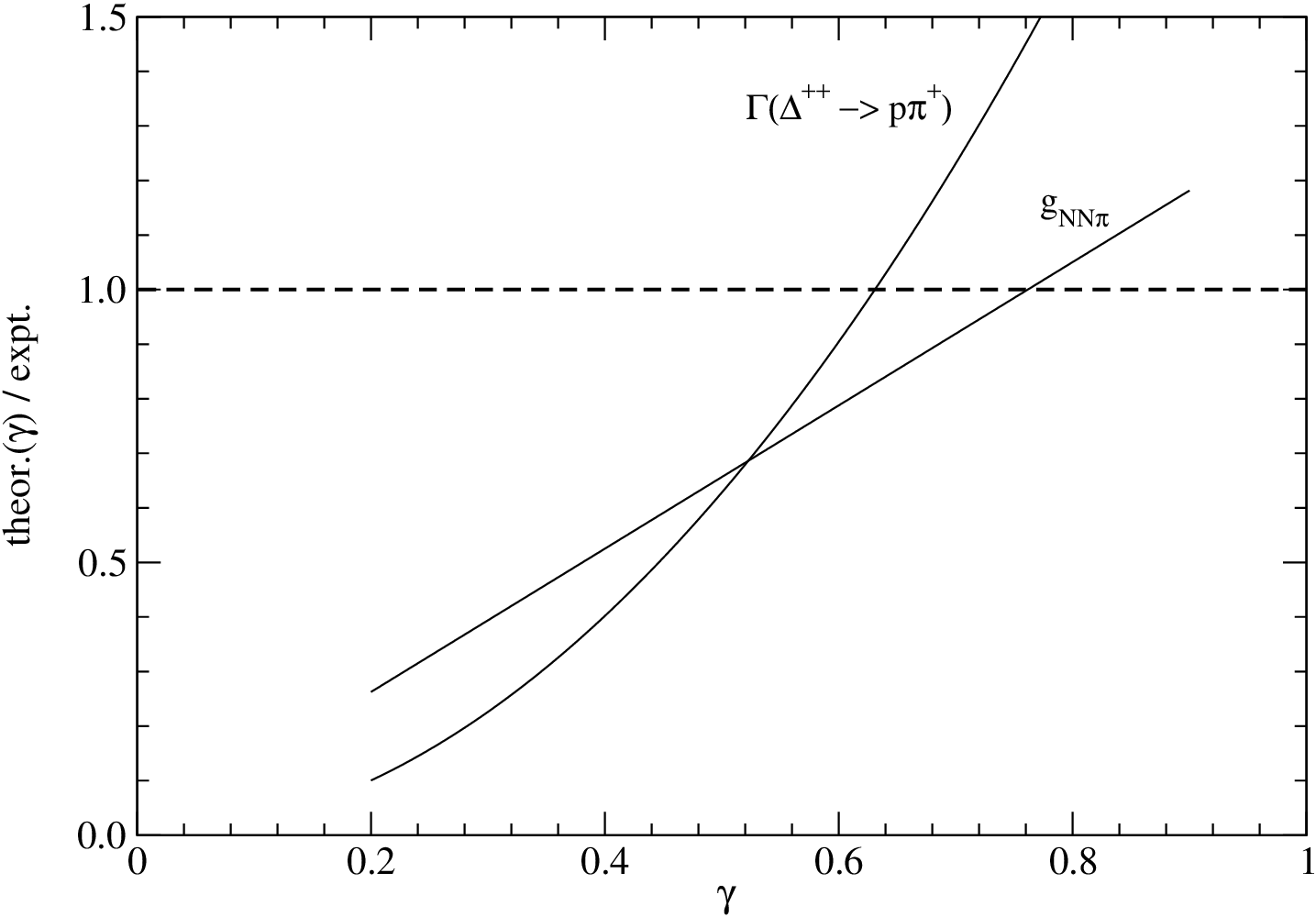}
\caption{
The ratio of theory and experiment for $\Gamma(\Delta^{++} \to p \pi^+)$
and $g_{\N\N\pi}$ given our standard quark model wavefunction length scales
$\alpha = \beta = 0.4$~GeV, but with the pair production amplitude $\gamma$ variable.}
\label{fig:scan_gamma}
\end{figure}

Alternatively, since the dimensionless pair production amplitude $\gamma$ 
represents poorly understood nonperturbative physics, it may have a different strength 
in baryon decays than in meson decays. In Fig.\ref{fig:scan_gamma} 
we show the ratio of theory to experiment for 
$\Gamma(\Delta^{++} \to p \pi^+)$ and $g_{\N\N\pi}$ 
for our ``standard" quark model 
baryon and meson wavefunction length scales $\alpha = \beta = 0.4$~GeV, varying
the pair production amplitude $\gamma$. Evidently with these wavefunctions 
a value of $\gamma \approx 0.7$ is favored for pion emission from light baryons, 
whereas $\gamma \approx 0.4-0.5$ gives the best description
of light meson decays.     

Nucleon couplings to the other ground state pseudoscalars
($\eta$ and $\eta'$) are simply related to $g_{\N\N\pi}$, 
provided that we assume 
identical spatial wavefunctions and pure $q\bar q$ states. 
Given the effective lagrangian
\be
{\cal L}_{\N\N\eta(')} = -ig_{\N\N\eta(')} \bar \Psi \gamma_5 \Psi 
\, \phi_{\eta(')},
\ee
and taking the $|\eta\rangle$ and $|\eta'\rangle$ flavor states to be 
the maximally mixed linear combinations 
$(|n\bar n \rangle \pm |s\bar s \rangle)/\sqrt{2}$, 
the NN$\eta(')$ coupling constants and form factors are related to the NN$\pi$ 
results by
\be
g_{\N\N\eta(')} = \frac{3}{2^{1/2}\cdot 5} 
\Big(\frac{m_{\eta(')}}{m_\pi}\Big)^{1/2} \, g_{\N\N\pi}
\ee
If we use $g_{\N\N\pi} = 13.5$ as input,
this gives $g_{\N\N\eta} = 11.5$ and 
$g_{\N\N\eta'} = 15.3$. Although these appear to be rather large 
NN$\eta(')$ couplings, their effect on NN scattering 
is suppressed by the larger $\eta$ and $\eta'$ masses 
in propagators and in the $1/\sqrt{2E}$ 
external line normalizations.

\subsection{NN$\sigma$}

The NN$\sigma$ coupling may be the most important nucleon-meson  
coupling in all of nuclear physics. In meson exchange models 
the exchange of a light scalar I=0 ``sigma" meson
is held to be the dominant mechanism underlying the 
intermediate ranged attraction, which is responsible for 
nuclear binding.
The balance between this attraction and the short distance repulsion 
in the nuclear core determines the equilibrium density of
$n_0 \approx 0.16$~nucleons/fm$^3$ in bulk nuclear matter. 
(Although pions are lighter and hence longer-ranged, and 
the $g_{\N\N\pi}$ coupling constant is quite large, the fact that pions
are emitted in a relative P-wave suppresses their contribution 
to the near-threshold interactions of nucleons.) 

Although the intermediate ranged attraction plays a crucial role 
in nuclear physics, the $\sigma$ meson itself is 
a dubious concept in meson spectroscopy. In I=0 $\pi\pi$ 
S-wave scattering one sees a very broad positive (attractive) phase shift, 
which if interpreted as an $s$-channel resonance would imply a mass of 
$ca.$ 1~GeV and a comparably large width. There are arguments from the 
quark model against a $q\bar q$ state with these properties; for example,
an I=0 $0^{++}$ $n\bar n$ resonance ($n=u,d)$ at this mass would have a 
rather large two-photon width of $\Gamma_{\gamma\gamma} \approx 2$~keV, and 
no such state is evident in $\gamma\gamma \to \pi^0\pi^0$. 
There is instead evidence in this reaction for a moderately broad scalar 
enhancement near 1.3~GeV, with about the expected two-photon width 
\cite{Eidelman:2004wy}; this broad $f_0(\approx 1300)$ is often identified with the 
I=0 $0^{++}$ $n\bar n$ quark model state. 

The explicit $\sigma$ meson included in meson exchange models 
has been explained as a parametrization of ``correlated 
two-pion exchange", so that its fitted strong coupling to NN
and low mass need not correspond to the properties of a physical P-wave
$n\bar n$ meson. The picture of the ``sigma meson" as a strongly mixed 
$(n\bar n)$-$\pi\pi$ state is supported by the large coupling predicted 
between these channels in the $^3\P_0$ model; the analogous 
S-wave kaon system is discussed in Ref.\cite{vanBeveren:2005pk}.

We can test the plausibility of sigma meson exchange models
of the intermediate ranged NN attraction by calculating the NN$\sigma$ 
coupling for a pure $n\bar n$ $\sigma$ meson,
using the same techniques we applied above to the NN$\pi$ system. If the 
sigma is dominantly a physical $n\bar n$ scalar meson, we would expect 
approximate agreement between the calculated and fitted $g_{\N\N\sigma}$ coupling constants.
If the sigma is instead a parametrization of two-pion exchange, agreement 
between the theoretical and fitted coupling constants would be fortuitous.

The calculation of the NN$\sigma$ coupling differs from the NN$\pi$ case
through the meson spin, space and isospin wavefunctions and the 
effective lagrangian. We assume the form
\be
{\cal L}_{\N\N\sigma} = - g_{\N\N\sigma} \bar \Psi  \Psi 
\, \phi_{\sigma}.
\ee
In our quark model description the $|\sigma\rangle$ is the usual I=0 
$|n\bar n\rangle$ flavor state  
$(|u\bar u \rangle + |d\bar d \rangle)/\sqrt{2}$ 
times the $|\J,\L,\S\rangle = |0,1,1\rangle$ angular momentum state
\bd
\frac{1}{\sqrt{3}} 
\Big( 
  |1,+1\rangle |1,-1\rangle
- |1, 0\rangle |1, 0\rangle
+ |1,-1\rangle |1,+1\rangle
\Big)
\ed
where the basis states are $|\L,\L_z\rangle |\S,\S_z\rangle$. 
The P-wave momentum space $q\bar q$ wavefunctions are given 
in Ref.\cite{Ackleh:1996yt}. On equating the 
effective lagrangian and quark model 
$h_{fi} = \langle p\sigma |H |p\rangle$ matrix elements,
we find 
\be
g_{\N\N\sigma} =  \gamma \; 2^{5/2} 3^{1/2} \pi^{3/4}\;  
\frac{r^{5/2}}{(1+r^2/3)^{5/2}}\;  
\frac{m_{\sigma}^{1/2} }{\alpha^{1/2} }.
\label{eq:g_NNsigma}
\ee
(A normalization convention dependent overall phase of $(-1)$ in our result 
is suppressed here.) The $\N\N\sigma$ form factor
is the quadratic 
$(1 + [(1+r^2/4)/9(1+r^2/3)] (\P/\alpha)^2)$ times
the Gaussian found in the $\N\N\pi$ case in Eq.(\ref{eq:NNpi_FF}). 

\begin{figure}[ht]
\vskip 0.7cm
\includegraphics[width=8.5cm]{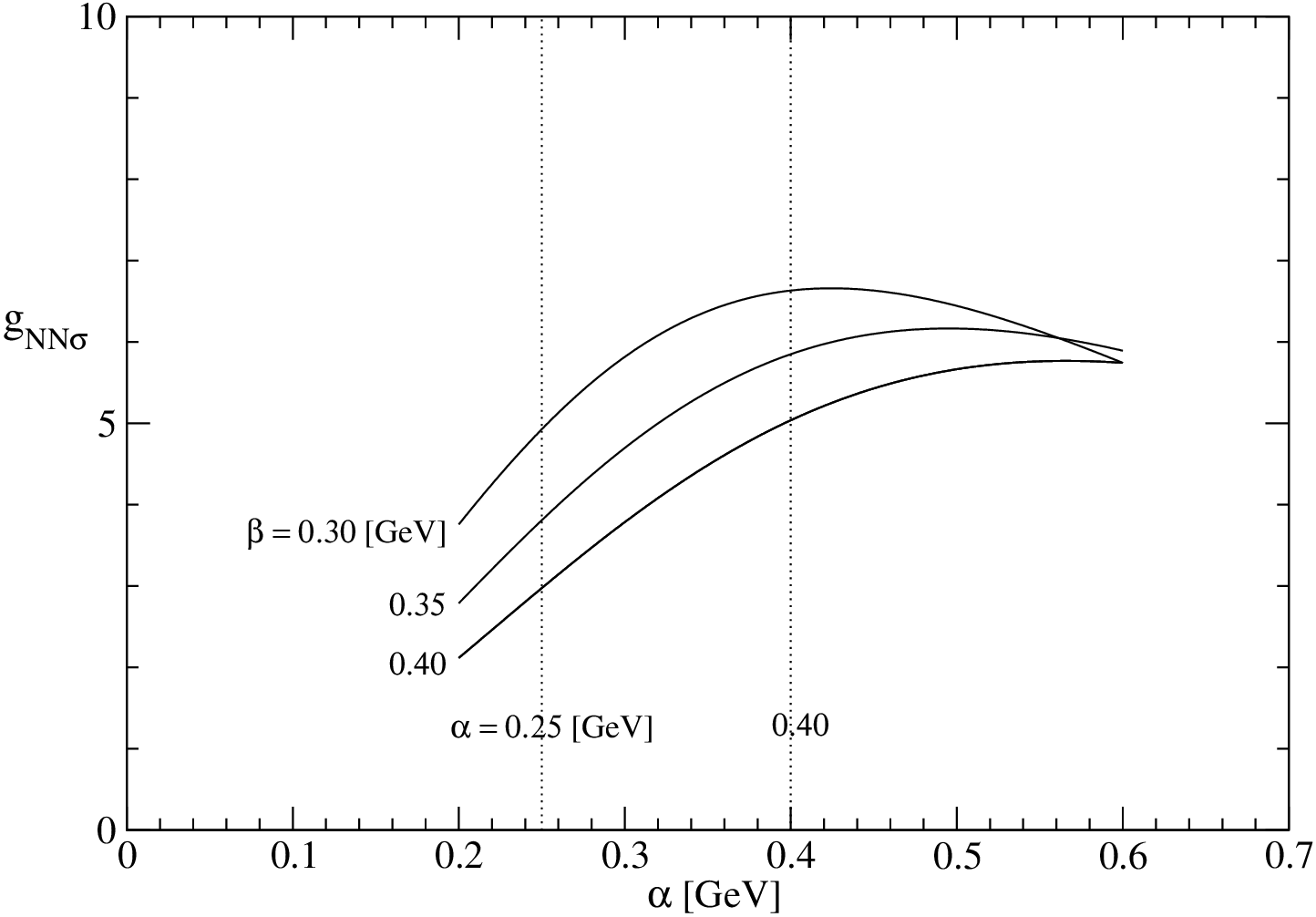}
\caption{
The theoretical nucleon - sigma meson coupling constant 
$g_{\N\N\sigma}$ 
(Eq.\ref{eq:g_NNsigma}).}
\label{fig:g_NNsigma}
\end{figure}

The numerical $g_{\N\N\sigma}$ predicted by Eq.(\ref{eq:g_NNsigma}) 
is shown in Fig.\ref{fig:g_NNsigma}
as a function of $\alpha$ and $\beta$ (assuming $m_{\sigma} = 500$~MeV). 
Evidently a value in the range $3-7$ is predicted by the quark model given
this $m_{\sigma}$,
with $g_{\N\N\sigma}\approx 5$ preferred.

Although it is of great interest to compare our calculated $g_{\N\N\sigma}$ coupling constant
with the values reported in meson exchange model fits to NN scattering data, there is unfortunately
no single consistent value reported for $g_{\N\N\sigma}$ in these models. 
The three best known meson-exchange models of the NN force 
in the literature are the
Paris \cite{Cottingham:1973wt,Lacombe:1980dr},
Nijmegen \cite{Nagels:1976xq,Nagels:1978sc,Stoks:1993tb}
and
Bonn \cite{Machleidt:2000ge} models, and their NN$m$ couplings are given in
Table~\ref{tab:ope_CCs}, together with our quark model results. The Paris model
did not consider a $\sigma$ meson.
In the recent ``CD-Bonn" model \cite{Machleidt:2000ge}, 
different $g_{\N\N\sigma}$ coupling constants and $\sigma$ masses are 
assumed in different NN channels; in the I=0 $^3$S$_1$ NN channel a 
$\sigma$ mass and coupling constant
of ${m_\sigma} = 350$~MeV and $g_{\N\N\sigma} \approx 2.5$ are used, whereas in I=0 $^1$S$_0$,
${m_\sigma} = 452$~MeV and $g_{\N\N\sigma} \approx 7.3$ are used. The $g_{\N\N\sigma}$ coupling is 
allowed to vary with L and I in the $\L > 0$ NN channels, and ranges from 1.9 to 9.9 
(see Table~XVI of Ref.\cite{Machleidt:2000ge}). In the Nijmegen model \cite{Nagels:1978sc}
a larger value of $g_{\N\N\sigma} = 17.9$ is quoted. Thus, although our quark model result 
$g_{\N\N\sigma} \approx 5$ is similar to the mean S-wave NN value quoted in the CD-Bonn 
model, the scatter in the fitted values of this parameter precludes
an accurate comparison between theory and experiment at present.
 
One may similarly evaluate the quark model prediction for the 
NN coupling of the $a_0$ I=1 partner of the $\sigma$. Given
the effective lagrangian
\be
{\cal L}_{\N\N a_0} = -g_{\N\N a_0} \bar \Psi \vec \tau\,\, \Psi 
\cdot \vec \phi_{a_0}
\ee
we find
\be
g_{\N\N a_0} = \frac{1}{3} 
\Big(\frac{m_{a_0}}{m_\sigma}\Big)^{1/2} \, g_{\N\N\sigma}.
\ee
Although I=1 scalar exchange contributes a somewhat smaller amplitude 
to NN scattering than I=0 exchange, it may nonetheless be possible to test
for the presence of both of these scalar meson exchange amplitudes through 
their interference, for example by comparing the I=0 and I=1 S-wave 
NN scattering amplitudes discussed above. 

\subsection{NN$\omega$ and NN$\rho$}

The NNV couplings are interesting in that the short-ranged repulsive core 
in the NN interaction has previously been attributed to vector meson exchange,
and the existence of the $\omega$ meson was regarded as support
for this picture. This mechanism now appears less plausible, since
the short range of vector meson exchange ($R\sim 1/m_{\omega} \sim 0.25$~fm)
implies extensive overlap of the NN quark wavefunctions.  

Evaluation of the NNV couplings and form factors uses the same procedure 
as the scalar and pseudoscalar couplings discussed above, although there
are complications due to the presence of two form factors and the non-transverse
components of the vector field.

As above we assume a term in the effective lagrangian for each coupling. 
For NN$\omega$ this lagrangian is
\be
{\cal L}_{\N\N\omega} = 
- g_{\N\N\omega}\, \bigg( 
\bar \Psi \gamma_\mu \Psi \, \omega_{\mu}
-
{\kappa_\omega \over 4\M_{\N}} \bar \Psi \sigma_{\mu\nu} \Psi \, \F_{\mu\nu}
\bigg)
\ee
where $\F_{\mu\nu} = \partial_\mu \omega_{\nu} -  \partial_\nu \omega_{\mu}$. 
We then equate
near-threshold Hamiltonian matrix elements $h_{fi}$ found from this effective 
lagrangian to the corresponding $^3\P_0$ decay model matrix elements. 
There is a complication in relating the relativistic effective lagrangian 
and nonrelativistic quark model matrix elements; we find that one must assume  
a vector meson polarization vector and four-momentum of the form 
$\epsilon_\mu = (0, \hat \epsilon)$ and $q_\mu = (0, \vec q\,)$ 
to equate these expressions.
The NN$\omega$ $\gamma_{\mu}$ and $\sigma_{\mu\nu}$ form factors 
may be separated by equating $h_{fi}$ matrix
elements with different spin states. The transitions
$p(+1/2) \to p(+1/2)\; \omega(0)$ and $p(+1/2) \to p(+1/2)\; \omega(-1)$ 
are useful in this regard, since they receive contributions from only 
the $\gamma_{\mu}$ and $\sigma_{\mu\nu}$ terms respectively. 
The resulting form factors are proportional to the NN$\pi$ result
Eq.(\ref{eq:g_NNpi}), since they involve the same Gaussian overlap integrals.
The NN$\omega$ coupling constants (and form factors) satisfy the relations
\be
g_{\N\N\omega}  = \frac{9}{5} 
\Big(\frac{m_{\omega}}{m_\pi}\Big)^{1/2} \, g_{\N\N\pi}
\label{eq:g_NNw}
\ee
and
\be
\kappa_{\omega}  = -3/2\ .
\label{eq:f_NNw}
\ee
The NN$\rho$ form factors, defined through the generalization of the
NN$\omega$ effective lagrangian to an I=1 $\rho$ meson,
\be
{\cal L}_{\N\N\rho} = 
- g_{\N\N\rho}\, \bigg(
\bar \Psi \gamma_\mu \vec \tau\, \Psi \cdot \vec{\rho}_{\mu}
-
\frac{\kappa_{\rho}}{4\M_{\N}} \bar \Psi \sigma_{\mu\nu} \vec \tau\, \Psi \cdot 
\vec {\F}_{\mu\nu}
\bigg)
\ee
(where 
$\vec {\F}_{\mu\nu} = 
\partial_\mu \vec \rho_{\nu} -  \partial_\nu \vec \rho_{\mu}$)
are related to the NN$\omega$ results by
\be
g_{\N\N\rho}  = \frac{1}{3}
\Big(\frac{m_{\rho}}{m_\omega}\Big)^{1/2} \, g_{\N\N\omega}
\label{eq:g_NNrho}
\ee
and
\be
\kappa_{\rho}  = +3/2\ .
\label{eq:f_NNrho}
\ee
The NN$\rho$ vector ($\gamma_{\mu}$) coupling constant was previously evaluated
by LeYaouanc {\it et al.} \cite{LeYaouanc:1972ae}. Their 
Eq.(3.13) for $g_{\N\N\rho}$ is consistent with our  
Eqs.(\ref{eq:g_NNpi},\ref{eq:g_NNrho}), 
provided that $i$) their factor of $3 R_{\N}^2 R_\rho^2 $ 
is actually $3 R_{\N}^2 R_\rho $ (their result as written is dimensionally incorrect), 
$ii$) their factor of $m_\rho^{3/2}$ should instead be $m_\rho^{1/2} m_{\N}$, 
analogous to their $g_{\N\N\pi}$ result, and 
$iii$) the factor of $\frac{1}{2}\vec \tau$ in their $\rho$ effective lagrangian 
Eq.(2.17) should be $\vec \tau$, as in their $\pi$ effective lagrangian 
Eq.(2.12).  LeYaouanc {\it et al.} did not evaluate the $\sigma_{\mu\nu}$ 
term, and did not consider the NN$\omega$ case.

It is interesting to compare our theoretical NN$V$ couplings with the 
fitted values found in meson exchange models of NN scattering.
If the $^3\P_0$ model is reasonably accurate in describing the coupling
between nonstrange baryons and vector mesons (which is currently being tested
at TJNAF in their search for missing baryon resonances decaying to 
N$\omega$ and N$\rho$), and the 
meson exchange models are correct in assuming that vector meson
exchange is the dominant mechanism underlying the short-ranged NN force,
we would expect to find approximate agreement between these couplings.

The fitted NNV couplings found in the three well-known meson exchange models
are given in Table~\ref{tab:ope_CCs}, together with our quark model results. 
Evidently we do not find good agreement.
Note in particular that the fitted strength of the dominant 
NN$\omega$ $\gamma_{\mu}$ coupling in the meson exchange models 
is about a factor of 2 smaller than the quark model prediction. 
The ratio of the NN$\rho$ to NN$\omega$ $\gamma_{\mu}$ couplings is rather 
similar in the two approaches; the meson exchange models 
quote a ratio of $\approx 0.2 - 0.4$, whereas the 
theoretical ratio (an SU(6) symmetry factor rather than a detailed
test of the quark model predictions) is $+1/3$.  
Although the ratio ``$\kappa$" of magnetic $(\sigma_{\mu\nu})$ to vector $(\gamma_{\mu})$
couplings does not yet appear to be well determined for both vectors in the meson exchange
model fits, there does appear to be agreement that 
$|\kappa_{\rho}| >> |\kappa_{\omega}|$. 
This is inconsistent with our quark model prediction
of equal magnitudes for these NNV ``strong magnetic" couplings, 
$\kappa_{(\omega,\rho)} = (-,+)\, 3/2$.
Since these ratios are simple SU(6) factors and do not involve 
uncertainties in the spatial wavefunctions, this disagreement may imply that vector meson
exchange is not the dominant short ranged NN interaction mechanism. This will be addressed 
in detail in a future study of the NN scattering amplitudes and phase shifts due to meson exchange,
augmented by quark model constraints on the nucleon-meson vertices.

\section{Summary and Conclusions}

In this paper we have developed a formalism for determining hadron 
strong vertices and form factors, ``three-point functions", 
in the context of the quark model. 
We apply this approach to the evaluation of meson-nucleon vertices,
several of which are important in meson exchange models of nuclear forces. 
The quark model expression we find for the NN$\pi$ coupling confirms 
an earlier Orsay result. With standard quark model parameters, 
this $g_{\N\N\pi}$  
about half the experimental value. 
Our quark model expression for the theoretical $g_{\N\N\sigma}$ 
strong coupling of nucleons to scalar mesons is a new result, 
and is numerically similar to the isospin-mean fitted NN S-wave value in the 
CD-Bonn model. Our quark model result for the NN$\rho$ vector ($\gamma_{\mu}$)  
coupling is consistent with an earlier Orsay result (after correcting 
typographical errors), although we find a nonzero magnetic ($\sigma_{\mu\nu}$) 
coupling. The strengths of the fitted NNV $\gamma_{\mu}$ couplings 
in meson exchange models are rather smaller than our quark model 
predictions. The NNV $\sigma_{\mu\nu}$ couplings are also not 
in good agreement with quark model predictions, although they 
may not be well determined in the current fits to NN scattering data. 

In future we plan to carry out calculations of the NN phase shifts predicted 
by meson exchange models, assuming quark model constraints on the NN$m$ couplings and 
form factors as derived here. This should allow a determination of the sensitivity 
of the data to parameters such as the 
$g_{\N\N\omega} /g_{\N\N\rho}$
and
$\kappa_{\omega} /\kappa_{\rho} $ ratios, 
for which we have definite quark model predictions. 
\vskip 0.3cm\phantom{.}

\acknowledgments

We acknowledge useful discussions with and communications
from S.Capstick, F.E.Close, A.Kerman, O.P\`ene, C.Thomas and H.V.von~Geramb.
This research was supported in part by the U.S. National Science
Foundation through grant NSF-PHY-0244786 at the University of Tennessee,
and by the U.S. Department of Energy through the 
Scientific Discovery through Advanced Computing Program (SciDAC) 
and through contracts
DE-AC05-00OR22725 at Oak Ridge National Laboratory and
DE-FG02-00ER41135 at the University of Pittsburgh.

%%%%%%%%%%%%%%%%%%%%%%%%%%%%%%%%%%%%%%%%%%%%%%%%%%%%%%%%%%%%%%%%%%%%%%%%%%%%%%

\begin{table*}
\caption{A summary of NN$m$ coupling constants. Our calculated 
values are shown in the middle columns, and fitted or assumed 
values in the meson exchange model literature are shown at right. 
Values in square brackets were fixed input.}
\vskip 0.5cm
\begin{tabular}{l|cc|ccc} 
\hline
Coupling 
& 
This ref.\footnote{Assumes ``standard" quark model parameters $\alpha = \beta = 0.4$~GeV
and $\gamma = 0.4$ (see text).}
\hskip 1cm  
& 
This ref.\footnote{Assumes $g_{\N\N\pi} = 13.5$.}
&
Paris \cite{Cottingham:1973wt,Lacombe:1980dr}
& 
Nijmegen \cite{Nagels:1978sc}
&
Bonn \cite{Machleidt:2000ge}
\\
\hline
\hline
$g_{\N\N\pi}$
&  7.1     
&  [13.5]  
&  [13.3]  
&   13.3   
&  [13.1]  
\\
\hline
$g_{\N\N\eta}$
&   6.0
&  11.5
&   -
&   9.8
&   -
\\
\hline
$g_{\N\N\eta'}$
&   7.9
&  15.3
&   -
&  10.5
&   -
\\
\hline
\hline
$g_{\N\N\sigma(\sim 500)}$
& 5.0
&  
&  -
& 17.9
& (2.5;\ 7.3)\footnote{This ``CD-Bonn" model introduces different $g_{\N\N\sigma}$ coupling constants 
for (I=0; I=1) NN channels, which would not be expected for an isosinglet $\sigma$.
In addition these $g_{\N\N\sigma}$ couplings and the $\sigma$ mass are allowed 
to vary with L (S-wave is quoted here), and a higher-mass $\sigma$ with large 
$g_{\N\N\sigma}$ couplings is also assumed.}
\\
\hline
$g_{\N\N a_0(\sim 1300)}$
&  2.7 
&  
&  -
&  3.3
&  -
\\
\hline
\hline
$g_{\N\N\omega}$ $(\gamma_{\mu})$
&  30.2  
&  
&  12.2
&  12.5
&  15.9
\\
\hline
$g_{\N\N\rho} / g_{\N\N\omega}$ $(\gamma_{\mu})$ 
& +1/3\footnote{Assumes $m_\rho = m_\omega$.}  
& 
& \hskip 1.5mm 0.43\footnote{This value is cited in Ref.\cite{Cottingham:1973wt} but is
not actually used in the Paris model, which does not incorporate $\rho$ exchange.} 
& 0.22
& 0.20
\\
\hline
$\kappa_\omega$  $(\sigma_{\mu\nu}/\gamma_{\mu})$
&  \hskip -2mm $-3/2$
& 
&  \hskip -2.5mm $-0.12$
&   0.66
&   0
\\
\hline
$\kappa_\rho$   $(\sigma_{\mu\nu}/\gamma_{\mu})$
& \hskip -2mm $+3/2$
& 
&   -
&  6.6
&  6.1
\\
\hline
\hline
\end{tabular}
\vskip 0.5cm
\label{tab:ope_CCs} 
\end{table*}
\end{document}